\documentclass{article}
\usepackage{ismir,amsmath,cite,url}
\usepackage{graphicx}
\usepackage{color}
\usepackage{amssymb}

\title{Local Music Event Recommendation with Long Tail Artists}

\twoauthors
  {Douglas Turnbull} {Ithaca College \\ {\tt dturnbull@ithaca.edu}}
  {Luke Waldner} {Ithaca College \\ {\tt lukezim5@gmail.com}}

\begin{document}

\maketitle
\begin{abstract}
In this paper, we explore the task of local music event recommendation. Many local artists tend to be obscure long-tail artists with a small \emph{digital footprint}.  That is, it can be hard to find social tag and artist similarity information for many of the artists who are playing shows in the local music community. To address this problem, we explore using Latent Semantic Analysis (LSA) to embed artists and tags into a latent feature space and examine how well artists with small digital footprints are represented in this space. We find that only a relatively small digital footprint is needed to effectively model artist similarity. We also introduce the concept of a \emph{Music Event Graph} as a data structure that makes it easy and efficient to recommend events based on user-selected genre tags and popular artists. Finally, we conduct a small user study to explore the feasibility of our proposed system for event recommendation.     
\end{abstract}
\section{Introduction}\label{sec:introduction}

Suppose that you move to a new city and are interested in exploring the local music scene. Typically, you might pick up the arts section of the local newspaper or go online to find a community notice board. Either way you would likely come to a long listing of music events where each event description would provide a small amount of contextual information: the names of the artists, the name and location of the venue, the date and start time of the event, the price of the tickets, and perhaps a few genre labels or a sentence fragment that reflects the kind of music you would expert to hear at the event.

While this ``public list of events'' model has been successful at getting fans to music events for many decades, we can use modern recommender systems to make music event discovery more efficient and effective. For example, companies like BandsInTown\footnote{https://www.bandsintown.com} and SongKick\footnote{https://www.songkick.com/} help users \emph{track} artists so that that the user can be notified when a favorite artist will be playing nearby. They also recommend upcoming events with artists who are similar to one or more of the artists that the user has selected to track. These services have been successful in growing both the number of users and in the number of artists and events covered by their service. For example, BandsInTown claims to have 38 million users and lists events for over 430,000 artists\footnote{According to https://en.wikipedia.org/wiki/Bandsintown on March 28, 2018.}. Event listings are added by aggregating information of ticket sellers (e.g., Ticketmaster\footnote{https://www.ticketmaster.com/}, TicketFly\footnote{https://www.ticketfly.com/}) and by artist managers and booking agents who have the ability to directly upload tour dates for their touring artists to these services.   

While this coverage is impressive, a large percentage of the events found in local newspapers are not listed on these commercial music event recommendation services. Many talented artists play at small venues (e.g., neighborhood pubs, coffee shops, and DIY shows) and are often not represented by (pro-active, tech-savvy) managers. Yet many music fans enjoy the intimacy of a small venue and a personal connection with local artists and may have a hard time discovering these events. 

As such, our goal is to develop a locally-focused music event recommendation system to help foster music discovery within a local music community. Here we define \emph{local} as all music events within a small geographic region (e.g., 10 square miles). This includes national and regional touring acts who may pass through town but it also includes non-touring artists (e.g.,  a high school punk band, a barber shop quartet, a jazz trio from the nearby music conservatory, or a neighborhood hip hop collective.)

What makes this problem technically challenging is that a large percentage of our local artists have a small \emph{digital footprint} or no digital footprint at all. That is, we may not be able to find these artists on sites that typically provide useful music information \cite{turnbull2008five} (e.g., Spotify\footnote{https://developer.spotify.com/}, Last.fm\footnote{https://www.last.fm/api}, AllMusic\footnote{https://www.allmusic.com/}). Similarly, we often do not have music recordings from these artists so we will not be able to make use of content-based methods for automatic tagging \cite{turnbull2008semantic} or acoustic similarity\cite{mcfee2012learning}. Rather, we will rely the small amount of contextual information that can be scraped from the event listings in the local newspaper or community notice board. 

We will first introduce the concept of a \emph{Music Event Graph} as a 4-partite graph that connects genre tags to popular artists to event artists to events. We then use latent semantic analysis (LSA) \cite{deerwester1990indexing} to embedding tags and artists into a latent feature space. We show that LSA is particularly advantageous when considering new or not well-known (long-tail) artists who have small digital footprints. This approach also allows us to independently control the \emph{popularity bias}\cite{turnbull2008five} of our event recommendation algorithm so that events with popular artists are no more or less likely to be recommended than events featuring more obscure local artists.

\section{Related Work}\label{sec:related-work}

We have been unable to find previous research on the specific task of music event recommendation though there is a significant amount of work on both music recommendation \cite{celma2010, schedl2017} (i.e., recommending songs and artists) and event recommendation \cite{macedo15,dooms11, minkov2010} (i.e., events posted on social networking sites.) It both cases, it is common to explore content-based (i.e., the substance of an item), collaborative filtering-based (e.g., usage patterns from a large group of users), and hybrid approaches. We  consider our approach to be a hybrid approach since we make use of both social tags (content) and artist similarity (collaborative filtering\footnote{While the details of the Last.fm algorithm for computing artist similarity remain a corporate trade secret, it would be reasonable to expect that these scores are computed using some form of collaborative filtering based on the large quantities of user listening histories that they collect \cite{barrington2009}.}) scores from Last.fm. 

As with many successful recommender systems, we make use of matrix factorization to embed data into a low dimensional space \cite{koren2009}. In particular, we use Latent Semantic Analysis (LSA) \cite{deerwester1990indexing} which is a common approach used in both text information retrieval \cite{manning2008} and music information retrieval systems (e.g., \cite{levy2007semantic, laurier2009, oramas2015semantic}). LSA is relatively easy to implement\footnote{For example, see http://scikit-learn.org/stable/modules/generated/ sklearn.decomposition.TruncatedSVD.html}, can improve recommendation accuracy, provides a compact representation of the data, works well with sparse input data, and can help alleviate problems caused by synonymy and polysemy \cite{manning2008}. We should note that other embedding techniques, such as probabilistic LSA \cite{hofmann1999probabilistic} and latent Dirichlet allocation (LDA) \cite{blei2003latent}, could also be used as an alternative to LSA.      






\section{Event Recommendation}


When developing an event recommendation system, we will consider an interactive experience with three steps:

\begin{enumerate}
\item \textbf{User selects genre tags}: Ask the user to select one or more tags from a list of board genres (``rock'', ``hip hop'', ``reggae'') based on the most common genres of the artists who are playing at upcoming local events.
\item  \textbf{User selects preferred popular artists}: Ask the user to select one or more artists from a list of recognizable mainstream artists (The Beatles, Jay-Z, Bob Marley) based on the selected genres and related to the artists who are playing an upcoming event.
\item \textbf{Display of recommended event list}: Show recommended events (with justification) to the user based on the the selected genre tag and popular artist preferences. 
\end{enumerate}
This is a common \emph{onboarding} process for both commercial music event services (e.g., BandsInTown) and music streaming services (e.g., Apple Music) since it quickly gives recommender systems a small but sufficient amount of music preference information for new users. After onboarding, a user can drill down into specific artists or events, as well as  listen to related music, explore a map of venues, etc.  

In this section, we describe the concept of a Music Event Graph and show how we can use it to efficiently recommend local music events based on the music preference information that is collected during user onboarding. 

\subsection{Music Event Graphs}

\begin{figure} 
\centering
  \includegraphics[width=3.3in]{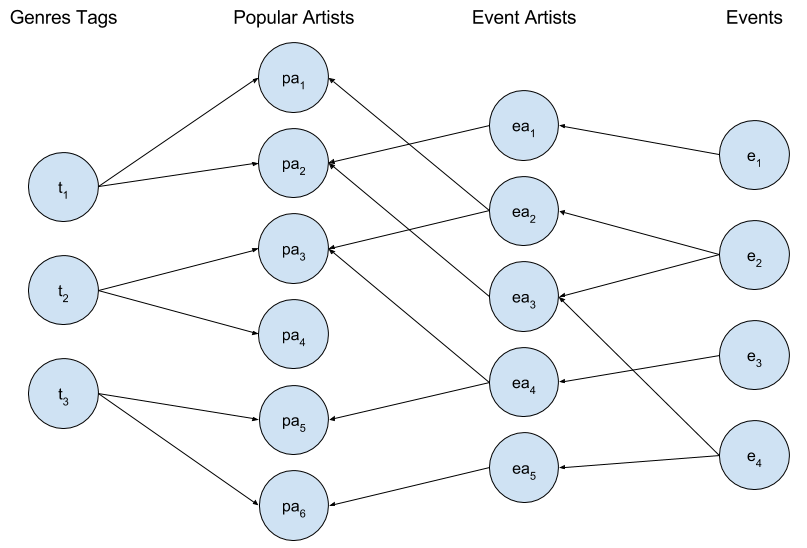}
  \caption{Construction a Music Event Graph:  First we collect \emph{events} and the \emph{event artists} who will be performing at these events (right nodes and edges). Then, we select \emph{genre tags} and related \emph{popular artists} (left nodes and edges). Finally, we connect event artists to popular artists based on our artist similarity calculation (middle edges).}
\label{fig:event-graph-creation}
\end{figure}

When considering event recommendation, there are two phases that we need to consider: offline computation of relevance information for all upcoming events and real-time personalized event recommendation. We will use a Music Event Graph to help us structure our event recommendation system. The music event graph is a k-partite graph with $k=4$ levels. Our four levels represent common genre tags, popular artists, event artists, and events as is shown in Figure \ref{fig:event-graph-creation}

To construct the graph, we follow the following steps:
\begin{enumerate}
\item Collect a set of upcoming local \emph{events}
\item Construct the set of \emph{event artists} from all of the local events
\item  Find the most frequently used \emph{genre tags} (e.g., ``rock'', ``jazz'', ``hip hop'') associated with the event artists. 
\item Using the genre tags, create a set of \emph{popular artists} by selecting the most well-known artists that are strongly associated with each genre.
\item For each event artist, find the most similar artists from the set of popular artists. 
\end{enumerate}
In Section \ref{sec:artist-similarity}, we will describe how we use harvested tags and artist similarity information to compute similarity between pairs of artists, as well as between artists and tags. These similarities are represented as real-valued weights, and as such, the event graph contains weighted edges.   

\begin{figure} 
\centering
  \includegraphics[width=3.3in]{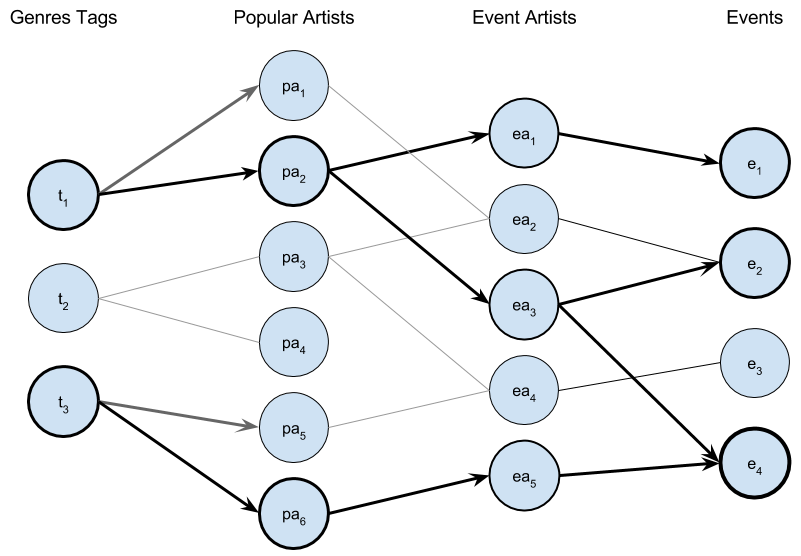}
  \caption{Construction the Music Event Graph: a user selects genre tags $\{t_1,t_3\}$. She is then shown popular artists $\{pa_1, pa_2, pa_5, pa_6\}$ and selects $\{pa_2, pa_6\}$. We then use the graph to  strongly recommend event $e_4$ with artists $ea_3$ and $ea_5$ based on multiple connections. We would also recommend events $e_1$ and $e_2$ based on their connections through $ea_1$ and $ea_3$, respectively. } 
\label{fig:event-graph-recommendation}
\end{figure}

Based on the interactive design described above, we can efficiently recommend events using a Music Event Graph. The user selects one or more preferred genres and then a set of relevant popular artists. Next our algorithm selects the event artists and their related events that are connected to the user's selected genres and popular artists. This graph traversal algorithm is depicted in Figure \ref{fig:event-graph-recommendation}. 

We note that our algorithm uses weighted edges to compute a user-specific relevance scores for each event as we move from left to right in the graph structure. In addition, we can use the graph structure to provide \emph{recommendation transparency} \cite{sinha2002role} by keeping track of the paths that are used to get from the user genre and popular artist selections to the recommend event artists and events. 

\section{Artist Similarity and Tag Affinity}\label{sec:artist-similarity}

At the core of the event recommendation system, we use Latent Semantic Analysis (LSA) when calculating artist similarity and artist-tag affinity. That is, we use truncated single value decomposition (SVD) to transform a large, sparse data matrix of artist similarity and tag information into a lower dimensional matrix such that each artist and tag is embedded into a dense, k-dimensional \emph{latent} feature space. Note that $k$ is a hyperparameter that is set based on empirical evaluation. We can then calculate artist-artist or artist-tag similarity using the cosine distance between pairs of vectors in this latent space.

Before we describe LSA, we will start with some useful notation for our problem setup:
\begin{itemize}

\item [$\mathcal{A}$]: set of artists.
\item [$\mathcal{T}$]: set of tags. Tags are any free text token that can be used to describe music. This may include genres, emotions, instruments, usages, etc. 
\item [$\mathcal{T}^G$]: a small subset of \emph{genre} tags (e.g., ``rock'', ``country'', ``blues'') that are frequently used to categorize music.
\item [$\mathcal{A}^P$]: set of \emph{popular} artists where each artist in the set is none of the most recognizable artists associated with at least one of the genre tags $\mathcal{T}^G$.
\item [$\mathcal{E}$]: set of local music events
\item [$\mathcal{A}^{\mathcal{E}}$]: set of \emph{event} artists where each artist has one or more upcoming events in $\mathcal{E}$
\item [$\mathcal{F}$]: set of features where $\mathcal{F}= \mathcal{A} \cup \mathcal{T}$. That is, we will describe each artist $a$ as a (sparse) feature vector of artist similarity and tag affinity values in $\mathbb{R}^{|\mathcal{A}|+|\mathcal{T}|}$  
\item [$X$]: (sparse) raw data matrix. The dimension of $X \in \mathbb{R}^{|\mathcal{A}| \times |\mathcal{F}|}$ where the $x_{i,j} \in [0,1]$ represents the affinity between the $i$-th artist (a row) and $j$-th feature (column). A value of 0 represents either no affinity or unknown affinity.  Note that all artists are self-similar so that $x_{i,i} = 1$. In terms of practical implementation, we can construct $X$ by stacking our $|\mathcal{A}| \times |\mathcal{A}|$ artist similarity matrix next to our$ |\mathcal{A}| \times |\mathcal{T}|$ artist-tag affinity matrix.   
\end{itemize}
LSA uses the truncated SVD algorithm to decompose the raw data matrix $X$ as follows:
\begin{equation}
X \approx X_k = U_k \Sigma_k V_k^{T}
\end{equation}
such that the matrix $X_k$ is a rank-$k$ approximation of $X$, $U_k$ is an $|\mathcal{A}| \times k$ matrix, $\Sigma_k$ is a diagonal $k \times k$ matrix of singular values, and $V_k^{T}$ is a $k \times |\mathcal{F}|$ matrix.  
We will then project each artist and tag in a $k$ dimensional latent feature space:
\begin{equation}
X_{SVD} = \Sigma_k V_k^{T}
\end{equation}
where $X_{SVD} \in \mathbb{R}^{k \times |\mathcal{F}|}$ or equivalently $\mathbb{R}^{k \times (|\mathcal{A}|+|\mathcal{T}|)}$ by construction. That is, the first $|\mathcal{A}|$ columns of $X_{SVD}$ represent artists and the last $|\mathcal{T}|$ columns represent tags all embedded into the same $k$ dimensional space.
We can also embed a new artist with raw feature vector $\mathbf{x} \in \mathbb{R}^{1 \times |\mathcal{F}|}$ by computing  
\begin{equation}
\mathbf{x}_{SVD} = \mathbf{x} V_k \Sigma_k^{-1}.
\end{equation}
so that $\mathbf{x}_{SVD} \in \mathbb{R}^{1 \times k}$ is projected in the same latent feature space.

Finally, we can compute artist-artist, artist-tag, or tag-tag similarity in the embedded space by comparing their respective (column) vectors in $X_{SVD}$. For example, if we have two latent feature vectors $p$ and $q$, we can compute their cosine similarity:
\begin{equation}
\cos(p, q) = \frac{p \cdot q}{||p||~||q||}
\end{equation}
where $p$ and $q$ are $k$-dimensional vectors and $||x|| = \sqrt[]{\sum_{i=1}^{k} x_i^2}$ is the l2-norm of a vector $x$. One nice property of cosine similarity, is that it tends to remove popularity bias. That is, we normalize the feature vectors by their length (l2-norm) such that each artist (and tag) vector is the same length. Without length normalization, popular artists which tend to have a bigger digital footprint (resulting in a denser raw feature vector with a bigger l2-norm) tend to produce larger similarity scores on average than if we did not normalize by length. 


\section{Event and Artist Data}\label{sec:data}

The data for our experiments is constructed by scraping local events from both TicketFly\footnote{https://www.ticketfly.com scraped February 15, 2018.} and the web-based public event calendar from a local newspaper\footnote{Details omitted during anonymous review process.}. We collected a total of $|\mathcal{E}| = 96$ events with 66 events from TicketFly, 36 events from the local newspaper, and 6 overlapping events between both websites. These events produced a set $|\mathcal{A}^{\mathcal{E}}| = 154$ event artists. We are also able to download short biographies of almost all of the event artists for events obtained from Ticketfly. The local newspaper only provides us with 1 to 3 genre tags for about half of the events we obtained from their site. 

We then used the Last.fm API \footnote{https://www.last.fm/api} to collect music information (popularity, biography text, artist similarity scores, and tags affinity scores) for each of our event artists. We then use snowball sampling on the similar artists and obtain this same Last.fm music information. We continue sampling these non-event artists until we have a set of 10,000 artists (i.e., $|\mathcal{A}| = 10,000$.) 

We define our set of tags $\mathcal{T}$ as the 1585 tags which are associated with 20 or more artists. Our set of genre tags $\mathcal{T}^{\mathcal{G}}$ are the top 20 tags which are most frequently associated with our event artists $\mathcal{A}^{\mathcal{E}}$. These include tags like ``rock'', ``jazz'',  and ``reggae''. However, we manually prune tags which are obviously not genres like ``seen live'' and  ``favorites''.  Finally, for each artist, we concatenate all available biographies (Last.fm, TicketFly, local newspaper) and attempt to find each of our tags in the combined biography text. If a tag is found, we label the artist with that tag. This is especially important since otherwise, many of our event artists would not be labeled with  any tags. In the end, we have 977,270 artist similarities and 456,867 artist-tag affinities.



\section{Exploring Artist Similarity in the Long Tail}

The core of our local event recommendation algorithm is our artist similarity calculation based on Latent Semantic Analysis (LSA). In this section, we show that most local event artists are relatively obscure \emph{long-tail} artists and that they tend to have small digital footprints. We also explore the relationship between digital footprint size and the accuracy of our artist similarity calculation. 

\subsection{Long-tail Event Artists}

\begin{figure} 
\centering
  \includegraphics[width=3.3in]{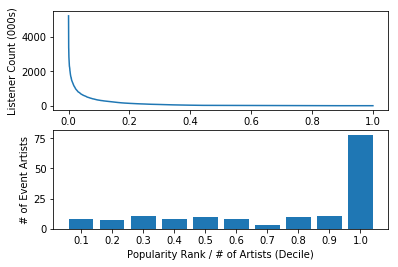}
  \caption{(Top) Plot of Last.fm Listener Count vs. popularity rank (divided by total number of artists) of 10,000 artists with most. (Bottom) Histogram of popularity for 154 event artists placed into 10 deciles of overall artist popularity. Note that most event artists reside in the long-tail of this popularity distribution.}
\label{fig:long-tail}
\end{figure}

In the top plot of Figure \ref{fig:long-tail}, we rank all 10,000 of our artists by their Last.fm listener counts. This shows a typical long-tail (power-law) distribution where a small number of popular artists in the short-head (left) receive much more attention than the vast majority of other artists in the long tail (right) \cite{celma2010,anderson2004long}.  
For example, 16.3\% of the most popular artists represent 80\% of the listener counts.  
In the bottom plot,  we show a histogram of the event artists' Last.fm listener counts broken down into deciles. We note that a disproportionate number of local event artists reside in the long-tale of this popularity distribution. In particular, 99 of the 154 event artists (64.2\%) are in the lowest three deciles of the ranking.

\subsection{The Digital Footprint of Event Artists }

\begin{figure} 
\centering
  \includegraphics[width=3.3in]{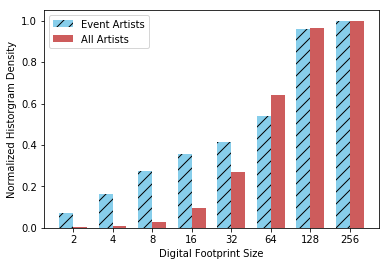}
  \caption{Cumulative distributions of digital footprint size (i.e., the number of nonzero artist similarity and tag affinity scores for each artists) for event artists and all artists. }
\label{fig:footprint-dist}
\end{figure}

As we discussed in the Introduction, obscure artists tend to have small digital footprints. To show this, we will consider the digital footprint of an artist to be the number of artist similarities plus the number of tag affinities for that artist. Equivalently, it is the number of nonzero values in the row of our raw data matrix $X$ that is associated with the artist. We note that digital footprint size is correlated with popularity rank ($r = -0.56$) such that popular artists tend to have a larger digital footprint. 

In Figure \ref{fig:footprint-dist}, we plot the empirical cumulative distribution for both event artists and all artists as a function of the digital footprint size. We see that about 27.2\% of the event artists have 15  or fewer digital footprints whereas only 2.8\% of all artists have so few digital footprints. This suggests that it will be important for us to design an artist similarity algorithm that works well in this \emph{small digital footprint} setting. 

\subsection{Artist Similarity with Latent Semantic Analysis}

\begin{figure} 
\centering
  \includegraphics[width=3.3in]{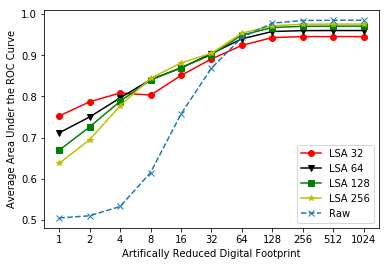}
  \caption{Plot of artist similarity ranking performance as a function of the  (artificially reduced) digital footprint size for various LSA embeddings with ranks of 32, 64, 128, and 256 dimensions. The \emph{Raw} approach represents computing cosine distances without first applying LSA to the raw artist data vectors. }
\label{fig:reduced-footprint}
\end{figure}

In Section \ref{sec:artist-similarity}, we introduced LSA as a algorithm for computing artist similarity. However, as we observed in the previous subsection, we are particularly interested in the case where an artist is represented by a small number of artist similarities and tag affinities (i.e., a small digital footprint.) To explore this, we will artificially reduce the digital footprint of artists to a fixed sized and see how well LSA is able to accurately compute artist similarity. 

To do this, we randomly split our data set of $\mathcal{A} = 10,000$ artists into a training set with $\mathcal{A}_{train} = 9,000$ artists and a test set of $\mathcal{A}_{test} = 1,000$ artists. Note that this involves removing 1000 rows \emph{and} 1000 columns from our raw data matrix $X$ since artists are also features. The training data will be used to to calculate our matrix decompositions $\mathcal{A}_{train} \approx U_k \Sigma_k V_k^{T}$ for a given embedding dimension $k$. 

Before projecting the $\mathcal{A}_{test}$ into the latent feature space, we limit the digital footprint size of each artists by randomly selecting artist similarity and tag affinity features to zero out.  We can then project $\mathcal{A}_{test}$ into the latent feature space and calculate the cosine distance between each pair of test set artists. Finally, we can calculate the Area Under the ROC Curve (AUC) \cite{manning2008} for each artist where the original artist similarities serve as the ground truth.

Figure \ref{fig:reduced-footprint} shows a plot of artificially reduced digital footprint size verses average AUC over the 1,000 test set artists for various LSA embedding dimensions. We also plot the curve for when we compute cosine distances between the \emph{raw} test artist vectors without projecting into a latent feature space. Here we note that LSA shows a improvement over raw cosine distance in small footprint setting of between 1 and 16 nonzero features. Once the digital footprint is larger than 128 nonzero features, the raw cosine approach slightly out performs LSA-based approach. However, the compactness of representing each artist with 32 or 64 floating point numbers may be advantageous in terms of storage size and computation time when we consider a much larger set of artists and tags. As such, we will use 64-dimensional LSA embeddings for the remaining experiments in this paper.    

\section{Exploring Event Recommendation}

To explore the performance of event recommendation using event graphs and LSA-based artist similarity, we conducted a small user study with a short 2-phase survey. We recruited 51 participants who were very familiar with the local music scene and attend live events in the area on a weekly basis. In the first phase of our survey, we asked participants to select between 1 and 3 genres from a set of 20 common genres. For each selected genre, the test subject was then asked to select between 1 and 3  artists from  a set of 16 popular artists that were representative of the genre (i.e., having a high cosine similarity score between the 64-dimensional latent feature vectors of the genre and the artist.)  In the second stage, participants were shown a list of the 154 event artists in our data set. They were asked to select all artists that they would like to see at a live event in the local area and were required to select 5 or more event artists.   

To evaluate our system, we use each test subject's selected genres or popular artists from phase 1 of the survey to rank order the 154 event artists using one of the approaches described below. In all cases, we embedded artists and tags into a 64-dimensional latent feature space using LSA with the data set that is described in Section \ref{sec:data}.  We then calculate the area under the ROC curve (AUC) for each user where ground truth relevance is determined from phase 2 of the survey.   

Each test subject provides multiple genre and multiple popular artist preferences. We explore a number of ways to combine these preferences to produce one ranking of the event artists for each test subject. We consider \emph{early fusion} and \emph{late fusion} steps for a number of approaches. In early fusion, we start with a set of latent feature vectors where each vector is associated with one of the users genre or artist preferences. We consider three approaches: 
\begin{itemize}
\item \textbf{average} the latent feature vectors into one vector
\item \textbf{cluster} the latent feature vectors and use the $k$ centroid vectors 
\item \textbf{none} use all of the latent feature vectors
\end{itemize} 
When clustering, we use the $k$-means clustering algorithm\footnote{http://scikit-learn.org/stable/modules/generated/ sklearn.cluster.KMeans.html} with the number of clusters ($k$) equal to the rounded natural log of the number of user preferences. 

For \emph{late fusion}, we must output one ranking of the event artists for each user. We consider three approaches
\begin{itemize}
\item \textbf{average cosine} ranks event artists by the average of the cosine similarity scores between the event artist vector and each vector in the set of user preference vectors.
\item \textbf{average rank} creates one ranking of event artists for each user preference vector, calculates the average rank for each event artist over this set of rankings, and then ranks them by this average rank.
\item \textbf{interleave} creates a set of rankings of the event artists for each user preference vector, and then constructs a final ranking by alternating between these ranking lists and picking top remaining artists that have not already been added to the final ranking.
\end{itemize}

\begin{table}[]
\centering
\caption{Event artist recommendation performance. The mean and standard deviation of AUC for our 8 expert test subjects when considering popular artist preferences, genre preferences, and both preferences together. See text for details on the seven approaches and the two baselines. }
\vspace{3mm}
\label{tab:event-artist-recommend}
\begin{tabular}{|l|c|c|c|}
\hline
\multicolumn{1}{|c|}{Approach} & \multicolumn{3}{|c|}{ User Preferences } \\
\hline
\multicolumn{1}{|c|}{Early / Late Fusion}   & Artists & Genres & Both \\
\hline
none / avg. cosine    & \textbf{.79} (.09) & .69 (.16) & .74 (.12) \\
none / avg. rank      & .75 (.11) & .66 (.15) & .76 (.09) \\
none / interleave     & \textbf{.79} (.11) & .69 (.15) & .71 (.15) \\
average / cosine      & \textbf{.79} (.09) & .69 (.16) & .74 (.12) \\
cluster / avg. cosine & .78 (.09) & .69 (.18) & .74 (.11) \\
cluster / avg. rank   & .74 (.13) & .66 (.20) & .68 (.17) \\
cluster / interleave  & .78 (.10) & .69 (.16) & .75 (.09) \\
\hline
\multicolumn{1}{|c|}{Baseline} & \multicolumn{3}{|c|}{~} \\
\hline
random & \multicolumn{3}{|c|}{.50 (.12)}\\
popularity & \multicolumn{3}{|c|}{.53 (-)}\\
\hline
\end{tabular}
\end{table}

Table \ref{tab:event-artist-recommend} shows average AUCs (and standard deviations) for our seven early/late fusion approaches when we use each user's popular artist preferences, genre tag preferences, and both sets of preferences together. We also include a popularity baseline that ranks all event artists by their Last.fm listener count as well as a random shuffle baseline.  We observe that artist preferences alone result in the best performance and a number of our proposed early/late fusion approaches produce similar results. 

We should also mention that we collected survey data from individuals who attended local shows on a less frequent (monthly) basis. The results for these test subjects was significantly lower (average AUC of 0.61) and more variable (AUC standard deviation of 0.15) for our best performing approach (Genre Preferences / None / Interleave.)  Having done error analysis on many of these less regular attendees, we often found that they selected a very eclectic set of event artists which did not match their preferences. As such, it would have been difficult for any recommender system to make accurate recommendations for many of these test subjects. This suggests that test subjects need to have a high level of familiarity with the local music community in order to provide useful ground truth for our experiment.

\section{Discussion}

In this paper, we explored the understudied task of local music event recommendation. This is an exciting task for the research community because it involves many interesting problems: long-tail recommendation,  the new user \& new artist cold start problems, multiple types of music information (artist similarity, tags), and user preference modeling. It is also an interesting problem outside of the academic research community since music event recommender systems can be used to help grow and support the local arts community. By promoting the work of talented local musicians, such systems can help fans discover new artists and help musicians reach new audiences. These audiences in turn attend more events which help sustain concert venues, music festivals, and other (local) businesses who benefit from direct ticket sales and other forms of indirect support (e.g., food, drinks, merchandise.)

While we were able to evaluate our system using a survey of local music experts, a more natural way to evaluate music event recommendation would be to build an interactive application thats collects user feedback over a longer period of time. We plan to develop such an app in the coming months and hope that it will be useful for expanding on the research that is presented in this paper.






\bibliography{Turnbull_ISMIR2018}

\begin{thebibliography}{10}

\bibitem{anderson2004long}
Chris Anderson.
\newblock The long tail.
\newblock {\em Wired magazine}, 12(10):170--177, 2004.

\bibitem{barrington2009}
Luke Barrington, Reid Oda, and Gert~RG Lanckriet.
\newblock Smarter than genius? human evaluation of music recommender systems.
\newblock In {\em ISMIR}, volume~9, pages 357--362. Citeseer, 2009.

\bibitem{blei2003latent}
David~M Blei, Andrew~Y Ng, and Michael~I Jordan.
\newblock Latent dirichlet allocation.
\newblock {\em Journal of machine Learning research}, 3(Jan):993--1022, 2003.

\bibitem{celma2010}
Oscar Celma.
\newblock Music recommendation.
\newblock In {\em Music recommendation and discovery}, pages 43--85. Springer,
  2010.

\bibitem{deerwester1990indexing}
Scott Deerwester, Susan~T Dumais, George~W Furnas, Thomas~K Landauer, and
  Richard Harshman.
\newblock Indexing by latent semantic analysis.
\newblock {\em Journal of the American society for information science},
  41(6):391, 1990.

\bibitem{dooms11}
Simon Dooms, Toon De~Pessemier, and Luc Martens.
\newblock A user-centric evaluation of recommender algorithms for an event
  recommendation system.
\newblock In {\em RecSys 2011 Workshop on Human Decision Making in Recommender
  Systems (Decisions@ RecSys' 11) and User-Centric Evaluation of Recommender
  Systems and Their Interfaces-2 (UCERSTI 2) affiliated with the 5th ACM
  Conference on Recommender Systems (RecSys 2011)}, pages 67--73. Ghent
  University, Department of Information technology, 2011.

\bibitem{hofmann1999probabilistic}
Thomas Hofmann.
\newblock Probabilistic latent semantic analysis.
\newblock In {\em Proceedings of the Fifteenth conference on Uncertainty in
  artificial intelligence}, pages 289--296. Morgan Kaufmann Publishers Inc.,
  1999.

\bibitem{koren2009}
Yehuda Koren, Robert Bell, and Chris Volinsky.
\newblock Matrix factorization techniques for recommender systems.
\newblock {\em Computer}, 42(8), 2009.

\bibitem{laurier2009}
Cyril Laurier, Mohamed Sordo, Joan Serra, and Perfecto Herrera.
\newblock Music mood representations from social tags.
\newblock In {\em ISMIR}, pages 381--386, 2009.

\bibitem{levy2007semantic}
Mark Levy and Mark Sandler.
\newblock A semantic space for music derived from social tags.
\newblock {\em International Society for Music Information Retrieval
  Conference}, 1:12, 2007.

\bibitem{macedo15}
Augusto~Q. Macedo, Leandro~B. Marinho, and Rodrygo~L.T. Santos.
\newblock Context-aware event recommendation in event-based social networks.
\newblock In {\em Proceedings of the 9th ACM Conference on Recommender
  Systems}, RecSys '15, pages 123--130, New York, NY, USA, 2015. ACM.

\bibitem{manning2008}
Christopher~D Manning, Prabhakar Raghavan, Hinrich Sch{\"u}tze, et~al.
\newblock {\em Introduction to information retrieval}, volume~1.
\newblock Cambridge university press Cambridge, 2008.

\bibitem{mcfee2012learning}
Brian McFee, Luke Barrington, and Gert Lanckriet.
\newblock Learning content similarity for music recommendation.
\newblock {\em IEEE transactions on audio, speech, and language processing},
  20(8):2207--2218, 2012.

\bibitem{minkov2010}
Einat Minkov, Ben Charrow, Jonathan Ledlie, Seth Teller, and Tommi Jaakkola.
\newblock Collaborative future event recommendation.
\newblock In {\em Proceedings of the 19th ACM international conference on
  Information and knowledge management}, pages 819--828. ACM, 2010.

\bibitem{oramas2015semantic}
Sergio Oramas, Mohamed Sordo, Luis~Espinosa Anke, and Xavier Serra.
\newblock A semantic-based approach for artist similarity.
\newblock In {\em ISMIR}, pages 100--106, 2015.

\bibitem{schedl2017}
Markus Schedl, Hamed Zamani, Ching-Wei Chen, Yashar Deldjoo, and Mehdi Elahi.
\newblock Current challenges and visions in music recommender systems research.
\newblock {\em arXiv preprint arXiv:1710.03208}, 2017.

\bibitem{sinha2002role}
Rashmi Sinha and Kirsten Swearingen.
\newblock The role of transparency in recommender systems.
\newblock In {\em CHI'02 extended abstracts on Human factors in computing
  systems}, pages 830--831. ACM, 2002.

\bibitem{turnbull2008five}
Douglas Turnbull, Luke Barrington, and Gert~RG Lanckriet.
\newblock Five approaches to collecting tags for music.
\newblock In {\em ISMIR}, volume~8, pages 225--230, 2008.

\bibitem{turnbull2008semantic}
Douglas Turnbull, Luke Barrington, David Torres, and Gert Lanckriet.
\newblock Semantic annotation and retrieval of music and sound effects.
\newblock {\em IEEE Transactions on Audio, Speech, and Language Processing},
  16(2):467--476, 2008.

\end{thebibliography}

\end{document}